\documentclass[letterpaper,twocolumn,10pt]{article}
\usepackage{usenix-2020-09}

\usepackage{tikz}
\usepackage{amsmath}
\usepackage{algorithm2e}
\usepackage{listings}
\usepackage[colorinlistoftodos]{todonotes}
\usepackage{float}

\usepackage{filecontents}
\usepackage{mlir}
\usepackage{hhline}
\usepackage{makecell}
\usepackage{multicol}
\usepackage{titling}
\usepackage{subfigure}
\usepackage{caption}
\usepackage{hyperref}
\usepackage{hhline}
\usepackage{multirow}
\usepackage{stfloats}
\usepackage{marvosym}

\newcommand{\loom}{{\textit{TileLoom}}}

\begin{document}
%-------------------------------------------------------------------------------

%don't want date printed
\date{}

\title{\Large \bf \textit{TileLoom}: Automatic Dataflow Planning for Tile-Based Languages on \\ Spatial Dataflow Accelerators}

\author{
Wei Li\textsuperscript{1,\Cross},
Zhenyu Bai\textsuperscript{1,\Cross,\Letter},
Heru Wang\textsuperscript{1, \Cross},
Pranav Dangi\textsuperscript{1,\Cross},\\
Zhiqiang Zhang\textsuperscript{1},
Cheng Tan\textsuperscript{2},
Huiying Lan\textsuperscript{3},
Weng-Fai Wong\textsuperscript{1},
Tulika Mitra\textsuperscript{1}
\\[0.5em]
\textsuperscript{1}School of Computing, National University of Singapore
\\
\textsuperscript{2}Arizona State University and Google, 
\textsuperscript{3}Lumai Ltd.
\\[0.5em]
\{liwei01, zhenyu.bai, heru.wang, dangi\}@nus.edu.sg
\\
t0937444@u.nus.edu, chengtan@asu.edu, huiying.lan93@gmail.com
{dcswwf,dcstm}@nus.edu.sg,
% \\[0.3em]
}

\maketitle

\begin{abstract}
Spatial dataflow accelerators are a promising direction for next-generation computer systems because they can reduce the memory bottlenecks of traditional von Neumann machines such as CPUs and GPUs. They organize computation around explicit, compiler-managed data movement over on-chip networks, allowing operands to be forwarded directly between processing elements and reducing reliance on high-latency, bandwidth-limited global shared memory. However, their performance depends strongly on how workloads are mapped to hardware. Naive mappings can perform poorly, and most users rely on hand-tuned vendor libraries. Thus, despite their potential for high performance, energy efficiency, and cost efficiency, limited programmability remains a major barrier to wider adoption.

This paper presents {\loom}, an MLIR-based end-to-end framework that compiles tile-based programs, such as Triton kernels, onto spatial dataflow architectures. Unlike compiler frameworks that focus on optimizing code generation \emph{within} a single tile, {\loom} distributes tile instances across spatially distributed cores and exploits the on-chip network and distributed memories to increase data reuse and reduce communication. {\loom} introduces a hardware representation that captures interconnect topology, memory hierarchy, and compute capabilities, enabling both architecture-specific optimizations and support for diverse spatial dataflow targets. In experiments on two generations of Tenstorrent systems, {\loom} achieves performance comparable to vendor libraries on various kernels.
\end{abstract}

\begingroup
\renewcommand{\thefootnote}{\Cross}
\footnotetext{These authors contributed equally.}
\endgroup

\begingroup
\renewcommand{\thefootnote}{\Letter}
\footnotetext{Zhenyu Bai is the corresponding author (\href{zhenyu.bai@nus.edu.sg}{zhenyu.bai@nus.edu.sg}).}
\endgroup

\section{Introduction}
Modern high-performance workloads, especially deep learning workloads, are highly data-intensive~\cite{data-intensive,ai-memory-wall}. For many of these workloads, the main bottleneck is not compute, but memory bandwidth~\cite{memory-wall-1, memory-wall-2, ai-memory-wall-2,memory-wall-tpu}. As process technology scales, we can place more arithmetic units on a chip, but off-chip memory bandwidth and capacity do not scale at the same rate~\cite{memory-scaling-1,memory-scaling-2}. Each DRAM access costs much more energy than an arithmetic operation~\cite{horowitz}. On-chip SRAM also becomes relatively more expensive in area and power as we move to smaller technology nodes~\cite{sram-scaling,horowitz}. Together, these trends make it hard for conventional, memory-centric von Neumann architectures such as CPUs and GPUs to keep their compute units busy.

\begin{figure}[t]
    \centering
    \includegraphics[width=\linewidth]{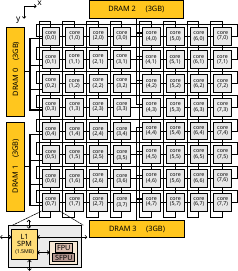}
    \caption{An example 2D-mesh spatial dataflow architecture, modeling Tenstorrent-Wormhole system.}
    \label{fig:2D-mesh}
\end{figure}

Spatial dataflow architectures are emerging as a strong alternative. Systems such as Tenstorrent~\cite{tenstorrent_wormhole}, Cerebras~\cite{cerebras_cs2}, Graphcore~\cite{graphcore_ipu}, SambaNova~\cite{sambanova_rda, sambanova_sn10}, Groq~\cite{groq_tsm, groq_tsp}, Meta's MTIA~\cite{mtia}, AWS's Trainium~\cite{trainium}, and Tesla's Dojo~\cite{tesla_dojo} organize computation around explicit, often software-controlled data movements over on-chip networks and buffers, reducing reliance on large shared caches and high-latency off-chip memories. When data is passed from core to core over short on-chip wires, the energy per bit and latency can be much lower than going back and forth to a shared cache or off-chip memory~\cite{horowitz,ho_future_wires,benini_noc}. Figure~\ref{fig:2D-mesh} shows a representative spatial dataflow accelerator architecture from Tenstorrent: a 2D grid of cores, each typically a SIMD or vector engine, often with a matrix unit, plus a local scratchpad, connected by a packet-switched mesh NoC. 64 cores can concurrently access their local scratchpads, yielding an aggregate peak bandwidth of roughly 24.5 TB/s—substantially higher than the 6 TB/s L2 bandwidth of the NVIDIA H100~\cite{thuning2024attention}. This abundant per-core bandwidth significantly alleviates memory bottlenecks for bandwidth-bound operators.

While the structure of spatial dataflow accelerators can provide high efficiency, it also creates a serious programmability problem~\cite{plasticine,graphcore_poplar,ipu_solvers}. Performance depends heavily on how the workload is mapped: which cores execute which parts of the computation, how data is partitioned across local memories, and how traffic is scheduled on the network~\cite{plasticine,q100,candles,indm,ditile_dgnn}. A naive mapping can cause severe load imbalance, network congestion, underutilized cores, or excessive off-chip memory traffic, resulting in poor performance and energy efficiency~\cite{q100,candles,indm,ditile_dgnn}. As a result, compiling high-level programs onto dataflow architectures is challenging, and non-experts struggle to write efficient low-level programs for them. Instead, users rely on vendor-provided, hand-optimized libraries by experts that implement a small set of popular kernels, such as matrix multiplication, convolutions, and attention, with carefully engineered mappings~\cite{cudnn,miopen,lewgpu,luthier,dsat}. This dependence limits both portability across architectures and the ability to experiment with new kernels or model structures~\cite{cudnn,luthier,dsat}.

A common abstraction for parallel accelerators is the grid–block–thread model popularized by CUDA’s grid-of-thread-blocks interface~\cite{cuda,kirk}. The programmer decomposes the problem into blocks and launches a grid of block instances across the device; each block, such as a CUDA thread block or OpenCL work-group, performs the same computation on a different region of the input or output tensor, and the grid covers the full problem domain~\cite{cuda,opencl}. Similar grid-based tiling models appear in high-level systems such as Halide and TVM~\cite{halide,tvm}, which treat tile shapes and launch configurations as schedule parameters and search over them to improve locality and parallelism.

On conventional, memory-centric architectures such as GPUs, the grid level is managed by hardware~\cite{cuda,kirk,nasa_gpu}: a hardware scheduler dynamically assigns blocks to Streaming Multiprocessors (SMs), and a shared cache hierarchy implicitly captures most data reuse across blocks~\cite{nasa_gpu,jia_volta}. This leaves little room for compilers to control grid-level placement, execution order, or communication. By contrast, the block–thread level is software-defined: the compiler or programmer decides how work within a block is mapped onto threads, warps, and intra-SM resources. CUDA exposes fine-grain control over thread organization, shared-memory layout, and synchronization, but exploiting it well requires careful tuning and detailed knowledge of GPU microarchitecture. To make this block-level programming more accessible, languages such as Triton~\cite{triton}, TileLang~\cite{tilelang}, CuTile~\cite{cutile}, Tilus~\cite{tilus}, and Taichi~\cite{taichi} let users express kernels in terms of high-level tile operators that define the work of a single block, while their compilers lower these operators onto intra-core or intra-SM resources.

Spatial dataflow accelerators change this picture. These architectures distribute compute and scratchpads across large arrays of processing elements connected by a high-bandwidth on-chip network, often without a large, unified cache that automatically exploits cross-core reuse~\cite{maeri,emerging_accels,mach_compiler}. Instead of offloading communication and placement decisions to hardware caches and schedulers, they expose richer inter-core programmability to software. Achieving good performance therefore requires a compiler or programmer to decide not only how to implement the tile program on a single core, but also how to place tile instances across cores and schedule them in time so that data can be forwarded or multicast efficiently over the NoC~\cite{maeri,timeloop,mach_compiler} and memory system. 
This inter-core programmability creates an enormous mapping design space: valid mappings are combinatorially large, and different mappings expose different reuse patterns and communication costs~\cite{timeloop,emerging_accels}. In today’s systems, these decisions are baked into vendor-specific compilers and libraries~\cite{sambanova_rda,emerging_accels,cerebras_sdk,mach_compiler,mlir-air,aries}, which encode architecture-specific strategies for placement, routing, and pipelining, but are time-consuming to develop and do not generalize easily to new kernels, models, or hardware generations.

By giving more control to software, spatial dataflow architectures are harder to program but can simplify hardware and improve efficiency. Crucially, \textbf{as the hardware becomes more transparent to software by exposing explicit cores, networks, and memories instead of opaque caches and schedulers, dataflow architectures become more predictable than memory-centric architectures.} With sufficient hardware information, \textbf{a compiler can more easily reason about the performance of different schedules and deduce good static mappings.}
Building on this insight, we propose \textbf{{\loom}, a compiler framework that supports end-to-end execution of tile-based programs on spatial dataflow systems.} Our goal is to let users write kernels in a tile-wise language, such as Triton, while the compiler maps the logical grid onto the physical array of cores, schedules tile execution in time, and orchestrates data movement and reuse over the NoC and distributed memories. In other words, {\loom} takes on responsibilities typically handled by hardware schedulers and runtimes on GPUs, moving these decisions to compile time for spatial dataflow systems. To support a broad range of dataflow architectures, {\loom} introduces a hardware description that models interconnect topology, memory hierarchy, and compute resources. Given this description, the compiler searches for mappings that balance load, improve data reuse, and respect network and memory constraints across architectures. We evaluate our approach on Tenstorrent systems and show that it can match or outperform vendor-provided handwritten libraries on key kernels.
\section{Framework}
\subsection{Overview}
\begin{figure}
    \centering
    \includegraphics[width=.85\linewidth]{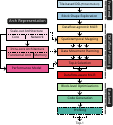}
    \caption{{\loom} framework overview.}
    \label{fig:compiler-stack}
\end{figure}

{\loom} compiles a kernel written in a tile-based DSL (currently supporting Triton and Helion) into an executable for a target spatial dataflow architecture.\footnote{The source code is available at \url{https://github.com/ecolab-nus/loom-dataflow}.} As shown in Figure~\ref{fig:compiler-stack}, the compiler stack is organized around three main components: a front-end that lower tile-level kernels into a standard dataflow-agnostic MLIR representation that we propose; a dataflow planning stage that decides spatiotemporal mappings, data movements and generate candidates in a standard dataflow-aware MLIR representation; a back-end that generates hardware-specific executables for each core, all guided by the multi-level architecture representation and performance model.

The \textbf{front-end} takes as input a tile-level kernel and a description of how that kernel is scaled out over the full problem (the launch grid). It explores candidate \emph{block shapes}: tile sizes and layouts, similar to the conventional auto-tunning process, and constructs corresponding programs. These candidate programs are then lowered into an MLIR-based intermediate representation and passed through normalization passes so that they share a common structure suitable for the dataflow planning pipeline. At this point, the computation is represented in a uniform, dataflow-agnostic IR affine + linalg + scf + arith): it encodes the tile and grid structure, but does not yet commit to any particular mapping onto the hardware.

The \textbf{dataflow planning} stage determines how this logical grid of tiles is realized on the target architecture. Guided by the hardware description, {\loom} explores spatiotemporal mappings of tile instances onto cores and time. For each candidate mapping, it performs data reuse analysis to identify which tiles can share data over the NoC or be reused over time via buffering, and it derives concrete data-movement plans: where each tensor tile is allocated, when it is copied, and how it is broadcasted or shared over the buffers. Together, the spatiotemporal mapping and the data-movement plan generate a design space of potential optimal dataflow planning candidates.

To model arbitrary dataflow architectures, the \textbf{architecture representation} is proposed, which provides the key inputs to both the dataflow planning search and the performance model. The scale-out (inter-core) description captures the spatial structure of the core array and interconnect, and guides spatiotemporal mapping decisions. The intra-core description captures the local memory hierarchy and compute resources, and guides decisions about where data is stored and how it is staged. {\loom} combines both levels of abstraction to generate \textbf{a performance model} that estimates the cost of different data-movement plans, taking into account memory bandwidths, NoC capabilities, and per-core compute throughput. The performance model is used to select the top candidates from the dataflow design space.

After dataflow planning stage, the IR is \emph{dataflow-aware}: memory allocations, copies, and communication endpoints are concretized according to the chosen mapping. After chosing the top candidates from the design space, the \textbf{block-level optimization and code generation} stage compiles the dataflow-aware program that runs on each core into the vendor’s existing back-end to generate an executable for each core. This corresponds to the existing compilers that compile tile-based DSLs to block-level program.

%These components together allow {\loom} to support a range of spatial dataflow architectures under varying levels of hardware detail. 
{\loom} uses an optional two-step selection strategy to provide better support for various architectures: the automatically generated performance model (using the architectural information) first ranks candidate dataflow plans and selects the top-$k$ mappings statically, and these $k$ candidates are then profiled on the real hardware to choose the final mapping (which bridge the small details not covered by the architecture representation). This combination of model-guided search and hardware validation enables {\loom} to produce high-quality mappings while remaining portable across different spatial dataflow systems, especially when micro-architectural details are missing and hence causing inaccuracy in the hardware modeling.

\subsection{Spatiotemporal Mapping}
\label{subsec:spatiotemporal}

\paragraph{Input representation.}
A unified dataflow-agnostic MLIR representation is required before the dataflow planning pipeline. We assume this representation produced by the front-end, we will cover the implementation details of the Triton front-end in Section~\ref{subsec:exp-setup}. An example of this MLIR for a matrix–multiplication kernel is shown in Listing~\ref{lst:mlir-before-dataflow}.
The block size and the strides for the input and output matrices are fixed by the front-end. The 2D output space is partitioned over two grid dimensions, \lstinline{x} and \lstinline{y}, which scale over the $m$ and $n$ dimensions of an $m \times k \times n$ matrix multiplication (an output-stationary tiling).

Scaling out across tiles is represented by an \lstinline{affine.parallel} loop over \lstinline{%block_id_x} and \lstinline{%block_id_y}. Inside this loop, an \lstinline{scf.for} loop iterates over the $k$ dimension and accumulates into the same output tile, representing the sequential execution within one block over one output tile. The front-end is required to "affinize" the address arithmetic for memory operations, so every load and store address is an affine function of the tile indices and intra-tile indices. The tile-wise computation itself is expressed with \lstinline{linalg} operations; this portion of the program is left unchanged during dataflow planning and is later lowered by the back-end.

The purpose of spatial–temporal mapping is to decide how the iteration space of the \lstinline{affine.parallel} loop—the logical multidimensional tile grid—is assigned to physical cores and to time. To preserve locality, {\loom} uses \emph{tiling-based} mappings: contiguous regions of the iteration space are mapped to contiguous spatial regions of the core array or to contiguous temporal regions in the execution schedule.

\begin{lstlisting}[float=*, style=mlir-druvbox-light-high-tiny, caption=MLIR representation before dataflow planning., label=lst:mlir-before-dataflow]
func.func @matmul(%A: memref<*xf32>, %B: memref<*xf32>, %C: memref<*xf32>, %grid_dim_x: index, %grid_dim_y: index, %grid_dim_z: index) {
    affine.parallel (%block_id_x, %block_id_y) = (0, 0) to (%grid_dim_x, %grid_dim_y) {
        %cst = arith.constant 0.000000e+00 : f32
        %0 = tensor.empty() : tensor<64x64xf32>
        %1 = linalg.fill ins(%cst : f32) outs(%0 : tensor<64x64xf32>) -> tensor<64x64xf32>
        %2 = scf.for %arg8 = 0 to 8 step 1 iter_args(%arg9 = %1) -> (tensor<64x64xf32>) {
            %4 = affine.apply affine_map<(d0, d1) -> (d1 * 32768 + d0 * 64)>(%arg8, %block_id_x)
            %reinterpret_cast_0 = memref.reinterpret_cast %arg0 to offset: [%4], sizes: [64, 64], strides: [512, 1] : memref<*xf32> to memref<64x64xf32, strided<[512, 1], offset: ?>>
            %alloc = memref.alloc() : memref<64x64xf32>
            memref.copy %reinterpret_cast_0, %alloc : memref<64x64xf32, strided<[512, 1], offset: ?>> to memref<64x64xf32>
            %5 = bufferization.to_tensor %alloc restrict writable : memref<64x64xf32> to tensor<64x64xf32>
            %6 = affine.apply affine_map<(d0, d1) -> (d1 * 32768 + d0 * 64)>(%block_id_y, %arg8)
            %reinterpret_cast_1 = memref.reinterpret_cast %arg1 to offset: [%6], sizes: [64, 64], strides: [512, 1] : memref<*xf32> to memref<64x64xf32, strided<[512, 1], offset: ?>>
            %alloc_2 = memref.alloc() : memref<64x64xf32>
            memref.copy %reinterpret_cast_1, %alloc_2 : memref<64x64xf32, strided<[512, 1], offset: ?>> to memref<64x64xf32>
            %7 = bufferization.to_tensor %alloc_2 restrict writable : memref<64x64xf32> to tensor<64x64xf32>
            // Tile-wise computation, omitted
            linalg.xxx;
            linalg.yyy;
            scf.yield %9 : tensor<64x64xf32>
        }
    }
}
\end{lstlisting}

\textbf{Schedule representation.}
On a 2D-mesh architecture like the one in Figure~\ref{fig:2D-mesh}, spatial–temporal mapping produces the loop structure shown in Listing~\ref{lst:after-spatiotemporal-mapping}. The outermost \lstinline{affine.parallel} loop now iterates over hardware spatial dimensions $x$ and $y$, each of size~8. These indices correspond directly to the cores in an $8 \times 8$ 2D mesh. After mapping, this loop represents code that actually runs in parallel across cores; its semantics have changed from a logical grid of parallelizable work-items to the physical parallel core indices.

The next \lstinline{affine.for} loops iterate over \lstinline{%tx} and \lstinline{%ty}. These loops enumerate \emph{waves} of tiles assigned to the same hardware array, i.e., temporal dimensions across blocks. Each wave assigns a batch of logical tiles to the available spatial cores, and the order of these waves determines the temporal schedule with which tiles traverse the array. The innermost \lstinline{scf.for} remains a purely sequential loop within each core’s program.\footnote{This sequential loop is also temporal, but we use “sequential’’ to distinguish intra-core sequencing from inter-tile temporal dimensions.}

\begin{lstlisting}[float=h, style=mlir-druvbox-light-high-tiny, caption=Loop structure after spatial--temporal mapping., label=lst:after-spatiotemporal-mapping]
affine.parallel (%x, %y) = (0, 0) to (8, 8) {
  affine.for %tx = 0 to %grid_dim_x ceildiv 8 {
    affine.for %ty = 0 to %grid_dim_y ceildiv 8 {
      scf.for %k {
        // tile-wise computation
      }
    }
  }
}
\end{lstlisting}

\textbf{Design space.}
The mapping from the original parallel dimensions to spatial and temporal dimensions defines the dataflow of the kernel. Under our tiling-based scheme, the design space is characterized by three coupled choices.

First, each original parallel dimension can be mapped to zero or more spatial dimensions. Mapping a parallel dimension to a spatial dimension corresponds to tiling the loop by the size of that spatial dimension and introducing a new outer \lstinline{affine.parallel} loop over the hardware index. %A parallel dimension that is not mapped to any spatial dimension remains purely temporal.

Second, when a parallel dimension is tiled by multiple spatial dimensions (for example, by both $x$ and $y$), the order in which tiling is applied matters. Different tiling orders induce different spatial layouts of tiles on the mesh and different execution schedules, and therefore expose different opportunities for spatial reuse and different communication costs.

Third, once all available spatial dimensions have been used, the remaining parallel dimensions become temporal dimensions implemented as loops over waves, (i.e. changing the rest of \lstinline|affine.parallel| to \lstinline|affine.for|), for instance the \lstinline{%tx} and \lstinline{%ty} loops in Listing~\ref{lst:after-spatiotemporal-mapping}. These temporal loops and their order determine how tiles are batched onto the array over time and in what order they revisit each region of the global iteration space, which in turn affects temporal reuse and the shape of NoC traffic.

{\loom} enumerates candidate spatial–temporal mappings by exploring combinations of these choices. Each mapping fixes a concrete loop nest structure, which the subsequent analyses use to reason about data placement, reuse, and communication.

\subsection{Data Reuse and Memory Operations}
\label{subsec:alloc-copy}

Different spatial–temporal mappings expose different opportunities to reuse data across time and across cores. {\loom} first analyzes these opportunities and then decides how to allocate data to memories and when and where to issue copies as broadcasts over the NoC or as loading from global memories. 
%For each load, as we have the access in the form of affine expression, we can easily determine along which spatial and temporal dimensions the accessed tile is invariant, and then choose whether to use the NoC (for spatial reuse) and at which loop level to place the load (for temporal reuse), subject to memory-capacity constraints.

\textbf{Reuse analysis on affine accesses.}
For a fixed spatial–temporal mapping (Section~\ref{subsec:spatiotemporal}), the loop nest contains: spatial loops (\lstinline{affine.parallel}) over hardware core indices; temporal loops (\lstinline{affine.for}) over waves of tiles, and sequential loops (\lstinline{scf.for}) inside each core, as shown in Listing~\ref{lst:after-spatiotemporal-mapping}.

The front-end expresses all memory accesses as affine functions of these loop indices. For each access, {\loom} inspects which induction variables appear in its affine expression. If an access does not depend on a spatial index such as \lstinline{%x}, then the accessed tile is identical for all cores along that dimension and is spatially reusable there. If an access does not depend on a temporal loop variable such as \lstinline{%tx}, then the same tile is used across all iterations of that temporal loop and is temporally reusable there. If the access depends only on sequential indices, then reuse is purely intra-core. {\loom} records this information as reuse annotations on the memory operations.

\textbf{Spatial reuse and broadcasts.}
We begin from a conservative baseline in which every core loads its tiles directly from global memory (for example, an L2 cache or DRAM) in the innermost loop, with no explicit sharing across cores.

If a load has no spatial reuse then the tile is unique to each core, so the load must remain a per-core global memory operation. If a load is spatially reusable along one or more spatial dimensions, {\loom} can reduce global traffic by replacing many per-core loads with a smaller number of global loads, followed by broadcasts over the NoC.

In the simplest case, if a tile is reusable only along a single spatial dimension, a designated producer core (or a small group of producers) loads it once from global memory and forwards it along that dimension (for example, along each row of the mesh), while receiving cores buffer their local copies. When a tile is reusable along multiple spatial dimensions, there are several concrete ways to realize that reuse. One option is to first duplicate the tile across all rows (or columns) and then perform independent one-dimensional broadcasts along the parallel dimensions; another is to propagate the tile in a wavefront-style pattern that sweeps across the array. These choices expose different tradeoffs between NoC traffic, latency, and local buffer usage. Listin~\ref{lst:spatial-reuse} shows one example candidate of the matrix multiplication mapped to the 2D-mesh example architecture (Figure~\ref{fig:2D-mesh} with a 2D dataflow where the A tiles are broadcasted for each row of cores through the horizontal links of the NoC and the B tiles for each column of cores through the vertical links (we will present the notion of network resources later in Section~\ref{subsec:hardware-representation}). The broadcast information is associated to the load instructions as annotations.

{\loom} does not fix a single strategy: it uses the network description from the hardware representation to enumerate the broadcast patterns that are legal, for each spatially reusable load, a small set of candidate implementations ranging from direct per-core global loads to one-dimensional and multi-dimensional broadcasts. {\loom} enumerate all the possibile combinations of all memory operations that creates a design space. Their different hardware costs will be taken into account later by the performance model to select the best ones.

 \begin{lstlisting}[style=mlir-druvbox-light-high-tiny,caption={Spatial Reuse},label=lst:spatial-reuse]
affine.parallel (x,y) = (0,0) to (8,8) {
  affine.for tm = 0 to M_waves {
    affine.for tn = 0 to N_waves {
      scf.for tk = 0 to K_tiles {
        load A[tm*8+x, tk]  {type="broadcast", resource={%noc_h}}
        load B[tk, tn*8+y]  {type="broadcast", resource={%noc_v}}
        // ...
      }
    }
  }
}
\end{lstlisting}

\textbf{Temporal reuse and loop hoisting.}
Temporal reuse is realized by choosing the loop level at which a load (or broadcast) is issued. At this point, the loop order is fixed by the spatial–temporal mapping; but we can hoist loads outward so that the same tile is reused across more iterations, at the cost of retaining it longer in a local buffer.

Consider the simplified GEMM-like loop nest below in Listing~\ref{lst:mlir-hoisting-both} left, treated as one candidate loop order, \lstinline{tm} $\rightarrow$ \lstinline{tn} $\rightarrow$ \lstinline{tk}:
the access \lstinline{A[tm, tk]} depends on \lstinline{tm} and \lstinline{tk}, but not on \lstinline{tn}, so tiles of \lstinline{A} are temporally reusable across the \lstinline{tn} loop. If we hoist the loads of \lstinline{A} outside the \lstinline{tk} loop, we must buffer all tiles \lstinline{A[tm, *]} for the current \lstinline{tm}. Because the address depends on \lstinline{tk}, hoisting across \lstinline{tk} enlarges the buffered region from a single tile \lstinline{A[tm, tk]} to the entire strip \lstinline|A[tm, 0..K_tiles-1]|. Hoisting further outward, above \lstinline{tn}, keeps the same buffered strip but reuses it across all values of \lstinline{tn}, as shown in Listing~\ref{lst:mlir-hoisting-both} right:

\begin{figure}[h]
\captionsetup{type=lstlisting} % use "Listing" instead of "Figure"
\centering
 \begin{minipage}[t]{0.48\linewidth}
 \begin{lstlisting}[style=mlir-druvbox-light-high-tiny]
affine.for tm = 0 to M_tiles {
  affine.for tn = 0 to N_tiles {
    scf.for tk = 0 to K_tiles {
      load A[tm, tk]  
      load B[tk, tn]  
      // ...
    }
  }
}
\end{lstlisting}
\end{minipage}
  \begin{minipage}[t]{0.48\linewidth}
 \begin{lstlisting}[style=mlir-druvbox-light-high-tiny]
affine.for tm = 0 to M_tiles {
  load A[tm, *]  
  affine.for tn = 0 to N_tiles {
    scf.for tk = 0 to K_tiles {
      load B[tk, tn]
      // ...
    }
  }
}
\end{lstlisting}
\end{minipage}
  \caption{Loop structures before and after hoisting.}
  \label{lst:mlir-hoisting-both}
\end{figure}

In general, hoisting a load across a loop that the access does not depend on increases reuse without increasing the size of the buffered region, because the accessed tile is the same for all iterations of that loop. Hoisting across a loop that the access does depend on expands the buffered region in proportion to the extent of that loop, because more distinct tiles must be kept live simultaneously. {\loom} applies these rules to enumerate, for each load or broadcast, all legal hoisting levels. For each level it computes the required buffer footprint and discards options whose footprint exceeds the capacity of the hardware model.

 \textbf{Temporal vs. Spatial.} Temporal reuse and spatial reuse are orthogonal. A tile can be reused only temporally (loaded once per core and reused across iterations), only spatially (broadcast once and immediately consumed), or in both ways (broadcast once and then reused across several temporal iterations). In all cases, the decision can be viewed as picking when a tile is first loaded or received and how long it remains live in local storage under a fixed loop order.

Combining these choices for all loads yields a concrete allocation and copy mapping: a description of which memory each tile resides in at each point in time and which NoC transfers occur. This schedule can be represented by a loop sctructure with annotations as shown a simplified example in Listing~\ref{lst:mlir-after-passes}. Each memory load is annotated with the target buffer, the type of load (using NoC for broadcasting or simple global load), and the NoC resources required. %The target buffer name and the NoC resource name are declared by the hardware representation language that we will present later in Section~\ref{subsec:hardware-representation}.
{\loom} prunes mappings that violate memory-capacity constraints and passes the remaining candidates to the performance model, which evaluates their compute, memory, and network costs and selects the top-$k$ mappings for back-end profiling.

\begin{lstlisting}[float=h, style=mlir-druvbox-light-high-tiny, caption=Example dataflow-friendly MLIR snippet before selection for matrix multiplication kernel on the example 2D-mesh architecture. , label=lst:mlir-after-passes]
affine.for tm = 0 to M_tiles {
  alloc A {target_buffer=%L1, size=K_tiles*block_M*block_K}
  load A[tm, *] {type="global", resources={%noc_h, %noc_v}}
  affine.for tn = 0 to N_tiles {
    alloc C {target_buffer=%L1, size=block_M*block_N}
    scf.for tk = 0 to K_tiles {
      alloc B {target_buffer=%L1, size=block_K*block_N}
      load B[tk, tn] {type="broadcast", resources={%noc_h, %noc_v}}
      // tile-wise computations
      load C
      linalg.matmul ...
      linalg.exp ...
      linalg.sqrt ...
    }
    store C {type="global". resources={%noc_h, %noc_v}}
  }
}
\end{lstlisting}

\subsection{Hardware Representation}
\label{subsec:hardware-representation}

{\loom} is designed to target multiple dataflow architectures. To make mapping decisions, the compiler needs a structured description of the hardware: how cores are arranged in space, where memories are placed, how components are connected, and what compute resources are available at each location. {\loom} captures this information in a multi-layer hardware representation stack. Different layers of this stack are consumed by different stages of the compiler passes.

We encode this representation in a custom MLIR dialect, \lstinline{df}. The dialect provides operators that describe the scale-out structure of the machine (cores and interconnects), the memory hierarchy and its connectivity, and the intra-core compute units. The performance model and the mapping passes process the program and the hardware description written this dialect together, rather than hard-coding any particular architecture.

\textbf{Scale-out architecture.} At the top level, the \lstinline{df} dialect describes the spatial layout of cores and the on-chip interconnect. The following operators are needed:

\lstinline{df.spatial_dim(size)} declares an abstract spatial dimension, used to index and replicate hardware components. Spatial dimensions naturally represent arrays of parallel resources such as cores or memories.

\lstinline{df.core(scaleout, scalein)} declares a set of cores indexed by the dimensions listed in \lstinline{scaleout}; the \lstinline|scalein| of the operation contains the compute components that live inside each core (optional argument, used at a lower abstraction level, described later).

\lstinline{df.interconnects(components, map, bandwidth)} declares a network (set of links) that connects a set of components according to an affine map \lstinline{map} and \lstinline{bandwidth} specifies the bandwidth per link. %Each instance of the map describes where traffic from a component at a given spatial index is sent, 

These operators are used to describe the scale-out architecture used in the spatial–temporal mapping. For our example 2D-mesh architecture (Figure~\ref{fig:2D-mesh}), we can describe it with:

\begin{lstlisting}[float=h, style=mlir-druvbox-light-high-tiny, caption=2D mesh with abstract scale-out., label=lst:df-2D-mesh]
%x = df.spatial_dim 8
%y = df.spatial_dim 8
%cores = df.core { scaleout = (%x, %y) }
%noc_h = df.interconnects %cores, %cores { map = affine_map<(d0, d1) -> ((d0 + 1) mod 8, d1)>,bandwidth = 28 }
%noc_v = df.interconnects %cores, %cores { map = affine_map<(d0, d1) -> (d0, (d1 + 1) mod 8)>, bandwidth = 28 }
\end{lstlisting}

This fragment describes an $8 \times 8$ array of cores connected by horizontal and vertical rings (or a torus). The modulo in the affine maps encodes wrap-around links. The spatial–temporal mapping pass uses this scale-out description to map the logical tile grid onto physical cores (Section~\ref{subsec:spatiotemporal}), and the performance model uses the \lstinline{df.interconnects} operators to estimate communication cost and traffic congestion of the memory operations (Section~\ref{subsec:perf-model}).

 \textbf{Memories and data movement.} When planning data movements (Section~\ref{subsec:alloc-copy}), {\loom} must reason about concrete buffers and how the network feeds data into them. To do so, the scale-out description is refined with explicit memories.

\lstinline{df.memory(scaleout, size, bandwidth)} declares a set of memories indexed by the \lstinline{scaleout} dimensions. Each instance has the given capacity and per-port bandwidth.

\lstinline{df.mux(dst, bandwidth, srcs, map)} declares a 1-to-\emph{N} connectivity between \lstinline{dst} components and \lstinline{srcs}, with topology specified by an affine \lstinline{map}. This operator captures fan-out connections such as “each core can access its local scratchpad” or “groups of cores share a DRAM channel”.

A lowered version of the 2D-mesh description that includes L1 memories and DRAM is shown in Listing~\ref{lst:df-2D-mesh-2}:

\begin{lstlisting}[float=h, style=mlir-druvbox-light-high-tiny, caption={2D mesh with scale-out cores, scratchpads, and DRAM.}, label=lst:df-2D-mesh-2]

%x = df.spatial_dim 8
%y = df.spatial_dim 8
%cores = df.core {scaleout=(%x, %y)}
// Per-core scratchpad memories.
%L1 = df.memory {scaleout=(%x, %y), size = 1499136, bandwidth = 60}
// Connect each core (x, y) to its local L1(x, y).
%core_to_L1 = df.mux %cores, %L1, {map = affine_map<(d0, d1) -> (d0, d1)>}
// On-chip NoC now connects L1 memories.
%noc_h = df.interconnects %L1, %L1, {map = affine_map<(d0, d1) -> ((d0 + 1) mod 8, d1)>, bandwidth = 28} 
%noc_v = df.interconnects %L1, %L1, {map = affine_map<(d0, d1) -> (d0, (d1 + 1) mod 8)>, bandwidth = 28} 
// Off-chip DRAM channels, indexed by a 1D spatial dimension.
%dram_idx = df.spatial_dim 4
%drams = df.memory {scaleout = %dram_idx, size = 12884901888, bandwidth = 267}
// Map each group of 4x4 edge cores to a DRAM channel.
%to_dram = df.interconnects %L1, %drams {map = affine_map<(d0, d1) -> (d0 ceildiv 4 + 2 * (d1 ceildiv 4))>, bandwidth = 30}
\end{lstlisting}

This representation now distinguishes the physical buffers that can hold tiles. For example, it specifies that each core \lstinline{(x, y)} has an L1 scratchpad of around 1.5\,MB, and that inter-core traffic flows between L1s rather than directly between cores. It also encodes DRAM channels are connected: every group of four adjacent cores along each edge shares the same DRAM bank, as in Figure~\ref{fig:2D-mesh}. When {\loom} chooses where to buffer a tile and the costs of memory operations, it uses these \lstinline{df.memory}, \lstinline{df.mux}, and \lstinline{df.interconnects} operators. As shown earlier in Listing~\ref{lst:mlir-after-passes}, load instructions are annotated with bindings to these physical resources.

\textbf{Intra-core compute model.}
To drive the performance model down to the level of individual cores, {\loom} needs a coarse description of the microarchitecture within each core. The \lstinline{df} dialect provides operators for this purpose:

\lstinline{df.mat(shape, throughput)} declares a matrix unit (for example, a tensor core) with a given input shape and sustained throughput.

\lstinline{df.vec(shape, throughput)} declares a vector unit with the given vector width and throughput.

\lstinline{df.scalar(latency)} declares a scalar unit with a given latency for scalar operations.

Each unit is assumed to accept operands of the specified shape and to produce results at the given throughput. These units are then attached to cores via the \lstinline{scalein} argument of \lstinline{df.core}, which describes the internal composition of each core. The lowest-level version of the 2D-mesh description therefore extends the previous listing with intra-core units:

\begin{lstlisting}[float=h, style=mlir-druvbox-light-high-tiny,
  caption=Extra specifications of the intra-core architecture, label=lst:df-2D-mesh-3]
%FPU = df.mat {shape=[32, 32, 32], throughput=98}
%SFPU = df.vec  {shape=[32], throughput=3}
%cores = df.core {scaleout = (%x, %y), scalein=(%FPU, %SFPU, [8, 1])}
\end{lstlisting}

With this information, the performance model can reason about the timing of both compute and memory. It can, for example, estimate how many cycles a particular tile-level matmul consumes on the matrix units and whether the NoC bandwidth and the L1 buffer are sufficient to keep them fed.

\begin{figure}
    \centering
    \includegraphics[width=\linewidth]{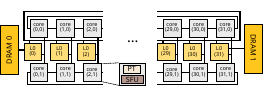}
    \caption{Example 1D triple-ring architecture modeled with the \lstinline{df} dialect.}
    \label{fig:1D-ring}
\end{figure}

\textbf{Expressiveness beyond 2D meshes.}
Although we have used the Tenstorrent-2D-mesh architecture as example, \lstinline|df| can describe other spatial dataflow architectures. For instance, a 1D triple-ring topology similar to the IBM-Spyre accelerator as shown in Figure~\ref{fig:1D-ring} can be described using \lstinline|df| program as shown in Listing~\ref{lst:df-1d-ring} .

\begin{lstlisting}[language=mlir, style=mlir-druvbox-light-high-tiny,
  float=h,
  caption=Example of the \lstinline{df} dialect describing a 1D triple-ring architecture.,
  label=lst:df-1d-ring]
module {
    // functional units
    %PT = df.mat {shape = [128, 128, 128], throughput=16384}
    %SFP = df.vec {shape = [128], throughput=128}
    // scale-out
    %x = df.spatial_dim 32
    %y = df.spatial_dim 2
    %cores = df.core "cores" {scaleout=(%x, %y) , scalein=(%PT, %SFP, [1,1])}
    %L1 = df.memory {scaleout=(%x) , size = 2097152, bandwidth = 128}
    %core_to_mem = df.mux %cores, %L1, {map = affine_map<(d0, d1) -> (d0)>}
    %small_rings = df.interconnects %cores, %cores, {map = affine_map<(d0, d1) -> ((d0 + 1) mod 8, d1)>, bandwidth = 32}
    %big_ring = df.interconnects %L1, %L1, {map = affine_map<(d0) -> ((d0 + 1) mod 8)>, bandwidth = 258}
    // Global (DRAM/L2)
    %d = df.spatial_dim 2
    %dram = df.memory {scaleout=(%d) , size = 34359738368, bandwidth = 512}
    %to_dram = df.interconnects %dram, %L1, {map = affine_map<(d0)->(d0*31)>}
}
\end{lstlisting}

\textbf{Discussion}
We structure the hardware representation as a stack of layers of abstractions because each compiler pass should depend only on the level of detail it actually needs. Spatial–temporal mapping requires only the scale-out structure of cores and the topology and bandwidth of the interconnect. Data-movement planning additionally needs to know where memories are placed and how they are wired to compute and to DRAM. The fine-grain performance model, in turn, needs a high-level view of intra-core compute units and their throughputs. This separation improves the reusability of the compiler. Changing the on-chip network, the memory hierarchy, or the per-core microarchitecture amounts to modifying the \lstinline{df} description, without rewriting the optimization passes and the rest of hardware description. It also creates a bridge from software-level mapping decisions to hardware-level design trade-offs. Starting from the lowest abstraction level, the same representation can be refined further to include implementation-specific costs such as area and power, enabling combined design space exploration over both mappings and hardware configurations—an important capability for spatial dataflow architectures, where the architecture itself varies widely across generations and vendors.

\subsection{Performance Modeling}
\label{subsec:perf-model}

After spatial–temporal mapping and data reuse / allocation decisions, {\loom} has a set of candidate dataflow schedules. Each candidate is represented as an MLIR program in which loop nests, memory operations, and data movements (global loads, broadcasts, buffered loads) are fully specified, and every memory or network operation is bound to concrete hardware resources described in the \lstinline{df} dialect (Section~\ref{subsec:hardware-representation}). A simplified example is shown in Listing~\ref{lst:mlir-after-passes}. The role of the performance model is to estimate the execution time of each candidate, using the compute units, memories, and interconnects defined in the \lstinline{df} description, and then select the top-$k$ candidates for downstream code generation and profiling.

\begin{figure}
    \centering
    \includegraphics[width=\linewidth]{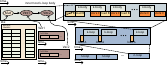}
    \caption{Pipelined execution of a matrix multiplication.}
    \label{fig:tiles-pipelin-exec}
\end{figure}

Figure~\ref{fig:tiles-pipelin-exec} illustrates how the performance model evaluates the overall execution time of the example in Listing~\ref{lst:mlir-after-passes}. It evaluates from the innermost loop outward, aggregating compute, memory, and network costs hierarchically.

\textbf{Compute cost per loop body.}
We first estimate the execution time of the innermost loop body, treating it as a block-level program running on a single core. For a given tile shape, every high-level operator (for example, a \lstinline{linalg.matmul}) is decomposed into the core’s low-level compute intrinsics. The available matrix, vector, and scalar units and their throughputs come from the \lstinline{df.mat}, \lstinline{df.vec}, and \lstinline{df.scalar} units attached to that core via \lstinline{df.core} (Section~\ref{subsec:hardware-representation}).

For each operator, {\loom} uses its \lstinline{linalg} semantics to recover the parallel iteration space. This tells us, for each functional-unit type, how many intrinsic invocations of that type are independent and can, in principle, be issued in parallel. We then conceptually schedule these intrinsics onto the available parallel units of the same type: if there are $N$ independent instances mapped to a unit type with $U$ identical units, each capable of issuing $r$ intrinsics per cycle, we approximate the operator’s time contribution on that unit type as $N / (U \cdot r)$ cycles.

{\loom} then accounts for data dependencies and resource sharing among different unit types. Operators that are independent and target different unit types (for example, a matrix multiply on a matrix unit and a pointwise activation on a vector unit) can execute in parallel, whereas dependent operators or operators that compete for the same unit type must execute in sequence. The loop-body compute time is approximated as the sum over sequential segments, where each segment’s time is the maximum over all operators that can run in parallel within that segment.

The potential parallelism exposed by this model is not necessarily fully achievable on a concrete microarchitecture, but the model does not attempt to exactly simulate the core’s instruction scheduler. Instead, it is calibrated to be accurate enough to distinguish compute-bound from memory-bound mappings and to reason about overlap between compute and data movement. In our experiments, this coarse-grain modeling of compute is sufficient to discriminate between different dataflow schedules.

\textbf{Compute–memory Overlap.}
Once we have an estimate for the loop body, we incorporate data movement. Let $T_{\text{load}}$ be the time spent on all loads in one iteration of the loop body, $T_{\text{compute}}$ the compute time (from the previous step), and $T_{\text{store}}$ the time spent on all stores. We assume that each iteration executes as a pipelined load–compute–store sequence with double buffering: while iteration $i$ is computing, the stores for iteration $i-1$ and the loads for iteration $i+1$ proceed in parallel whenever possible.

For an innermost loop with $I$ iterations, the total execution time is approximated as:
\begin{multline*}
  T_{\text{loop}} \approx (I - 2) \cdot
    \max\!\bigl(T_{\text{load}} + T_{\text{store}},\, T_{\text{compute}}\bigr) \\
  + \max\!\bigl(T_{\text{load}},\, T_{\text{compute}}\bigr)
  + \max\!\bigl(T_{\text{store}},\, T_{\text{compute}}\bigr)
  + T_{\text{load}} + T_{\text{store}}.
\end{multline*}
The first term accounts for the $I - 2$ steady-state iterations, where load, compute, and store can overlap and the throughput is limited by the slower of \mbox{$T_{\text{compute}}$} and $T_{\text{load}} + T_{\text{store}}$. The remaining terms account for filling and draining the pipeline. This behavior is illustrated in Figure~\ref{fig:tiles-pipelin-exec}, where the $k$-body is executed in parallel with loading the next tile of $B$ and storing the result tile of $C$ from the previous iteration.

\textbf{Concurrent data transfers and network traffic.}
Several memory operations may occur simultaneously and create traffic contention over the NoC. {\loom} estimates the effective bandwidth of each memory operation under this contention, using both the interconnect structure described in the hardware representation and the mapping annotations attached during memory operation mapping (Section~\ref{subsec:alloc-copy}). At that mapping step, each load or store is lowered either to a global load/store or to a broadcast pattern, and the compiler records which subsets of the network it uses. For global loads, we assume accesses are sufficiently random that traffic is spread across the NoC links. For broadcasts, the resources used depend on the chosen pattern: on the 2D-mesh example, a broadcast that is performed independently along each row uses only the horizontal ring links (such as \lstinline|%noc_h|), whereas a broadcast over the entire mesh may exercise both horizontal and vertical rings (both \lstinline|%noc_h| and \lstinline|%noc_v|). These choices are fixed during memory operation mapping and appear as annotations, as in Listings~\ref{lst:spatial-reuse} and~\ref{lst:mlir-after-passes}.
Given these annotations, the performance model groups memory operations according to the network links and memory interfaces they occupy. For each group of operations that share a particular subset of links, it aggregates their offered traffic and derives an effective bandwidth per transfer by partitioning the nominal link bandwidth among them. Equivalently, the bandwidth seen by any one operation is reduced in proportion to the number of concurrent transfers using the same links or banks. The transfer time for each load or store is then computed from its tile size and effective bandwidth. The per-iteration load and store times, $T_{\text{load}}$ and $T_{\text{store}}$, are obtained by combining these transfer times across operations, treating transfers on disjoint link sets as running in parallel and transfers on overlapping link sets as time-sharing the same resources. These $T_{\text{load}}$ and $T_{\text{store}}$ values are then plugged into the pipelined overlap model described above.

\textbf{Candidate ranking through auto-profiling.}
{\loom}'s optimizations are highly dependent on the accuracy of the architectural models. When the architectural and micro-architecral information is not accurate enough, {\loom} supports a profiling-based autotuning step to maintain its performance. Concretly, for each candidate schedule, {\loom} combines the compute and data-movement estimates with the architectural model to obtain an approximate end-to-end execution time. This time reflects the balance between compute and communication, the benefit of spatial and temporal reuse, and the impact of NoC and memory contention on the concrete hardware described by \lstinline{df}. 
{\loom} then ranks all candidates by this estimated time and keeps only the top-$k$ dataflow mappings. Only these top-$k$ candidates are handed to the back-end for full code generation and on-hardware profiling, where the final best-performing configuration (top-1) is selected. The choice of $k$ controls the trade-off between compile time and the likelihood of including the true optimum: a larger $k$ explores more mappings but costs more compilation and profiling time. We study this trade-off in Section~\ref{sec:ablation_topk}.
\section{Evaluation}
\subsection{Experimental Setup}
\label{subsec:exp-setup}

\textbf{Hardware platform.}
We conduct our experiments on Tenstorrent Wormhole and Blackhole cards. Table~\ref{tab:tt-specs} summarizes the relevant hardware specifications. We compile {\loom} programs on a host machine equipped with dual 16-core Intel Xeon Gold 6326 CPUs and 512~GB of DRAM.

\begin{table}[h]
    \centering
    \caption{Specifications of the Tenstorrent Wormhole and Blackhole used in our evaluation.}
    \label{tab:tt-specs}
    %\resizebox{.95\linewidth}{!}{
    \begin{tabular}{c|c|c}
    \hline
                        & TT-Wormhole   & TT-Blackhole\\
                        \hline
         Topology       & 8 $\times$ 8  & 12 $\times$ 10 \\
         On-chip SRAM   & 108 MB        & 180 MB \\
         DRAM           & 12 GB GDDR6   & 32 GB GDDR6 \\
         Off-chip bw.   & 288 GB/s      & 512 GB/s \\
         TFLOPS (FP16)  & 64            & 162 \\
         TBP            & 160 W         & 300 W \\
         Technology Node& GF 12nm~\cite{tt-wormhole-semianalysis} & TSMC 6nm~\cite{tt-blackhole-semianalysis} \\
        \hline
    \end{tabular}
    %}
\end{table}

\textbf{Architecture targets and modeling.}
Each Wormhole and Blackhole chip contains a 2D array of cores, with $8 \times 8$ cores on Wormhole and $12 \times 10$ cores on Blackhole, as shown in Figure~\ref{fig:2D-mesh}. We model the Tenstorrent architecture in the \lstinline{df} dialect, as described in Figure~\ref{fig:2D-mesh} and Section~\ref{subsec:hardware-representation}. Because the complete proprietary hardware specification is unavailable, we recover key parameters through isolated microbenchmarks, including matrix/vector-unit throughput and effective NoC and DRAM bandwidths. We instantiate these measured values in the \lstinline{df} hardware representation used by {\loom}'s performance model.

\textbf{Frontend.}
{\loom} currently supports Triton and Helion as kernel-development frontends. Although the core compiler is frontend-agnostic, our experiments use either Triton or Helion depending on kernel availability; the source code for all input kernels is available in our GitHub repository. For both frontends, we tune the tile, or block, shape using their existing Python-based autotuning stacks. For Triton, we use \textit{triton-shared}~\cite{triton-shared} to lower kernels into MLIR, then apply a custom affinization pass that rewrites index arithmetic into affine expressions, followed by normalization into our dataflow-agnostic MLIR format, as shown in Figure~\ref{fig:compiler-stack}. For Helion, after Python-level tuning, we lower Helion Device IR into the same standard MLIR format using our \href{https://link-to-be-given-once-public}{custom lowering tool}.

\textbf{Backend.}
We lower our dataflow-aware MLIR to TT-Metalium, Tenstorrent's low-level C API, to generate the final executable, as shown in Figure~\ref{fig:compiler-stack}. TT-Metalium exposes coarse-grained primitives for computation, synchronization, buffer allocation, and data movement, and handles most block- and core-level optimizations. {\loom} bridges the dataflow-level program and TT-Metalium by performing lifetime analysis over the block-level compute graph, using the resulting lifetimes to determine buffer allocation and synchronization among memory, compute, and data-movement operations. TT-Metalium then lowers these coarse-grained operations to hardware instructions.

% \subsection{End-to-End Performance}

% Table~\ref{tab:all-perf} reports the geometric mean of {\loom}'s relative performance over TTNN on four representative kernels: GEMM, FlashAttention, Flash Decode, and Mamba Chunk Scan linear attention. For each kernel, we evaluate a range of input shapes; detailed shape-by-shape results are discussed below. We report the performance of the top-ranked candidate selected by {\loom}'s performance model, without the final profiling-based tuning step, so the speedups come solely from {\loom}'s architectural modeling and dataflow-planning strategy.

% \begin{table}[h]
%     \centering
%     \caption{Relative performance of {\loom} over TTNN on GEMM, FlashAttention, Flash Decode, and Mamba Chunk Scan linear attention.}
%     \label{tab:all-perf}
%     %\resizebox{.8\linewidth}{!}{
%     \begin{tabular}{c|c|c}
%         \hline
%         {\loom} vs. TTNN & Wormhole & Blackhole \\
%         \hhline{---}
%         FlashAttention & 1.94x & 1.98x \\
%         Flash Decode & 0.84x     & 0.87x \\
%         Mamba Chunk Scan & 27.23x & 16.27x \\
%         GEMM & 0.95x & 1.10x \\
%         \hline
%     \end{tabular}
%     %}
% \end{table}
\subsection{End-to-End Performance}

Table~\ref{tab:all-perf} reports the geometric mean of the relative performance of {\loom} over TTNN on four representative kernels: GEMM, FlashAttention, Flash Decode, and Mamba Chunk Scan linear attention. For each kernel, we evaluate a range of input shapes. The detailed shape-by-shape results are discussed in the corresponding subsections below. In this section, we report the performance of the top-ranked candidate selected by {\loom}'s performance model, without the final profiling-based tuning step. Therefore, the reported speedups come solely from {\loom}'s architectural modeling and dataflow-planning strategy.

\begin{table}[h]
    \centering
    \caption{Relative performance of {\loom} over TTNN on GEMM, FlashAttention, Flash Decode, and Mamba Chunk Scan linear attention.}
    \label{tab:all-perf}
    %\resizebox{.8\linewidth}{!}{
    \begin{tabular}{c|c|c}
        \hline
        {\loom} vs. TTNN & Wormhole & Blackhole \\
        \hhline{---}
        \multicolumn{3}{c}{vs. library implementation} \\
        \hhline{---}
        FlashAttention & 1.94x & 1.98x \\
        Flash Decode & 0.84x     & 0.87x \\
        GEMM & 0.95x & 1.10x \\
        \hhline{---}
        \multicolumn{3}{c}{vs. unfused implementation} \\
        \hhline{---}
        Mamba Chunk Scan & 27.23x & 16.27x \\
        \hline
    \end{tabular}
    %}
\end{table}

\subsubsection{FlashAttention}

\begin{figure}[h]
    \centering
    \includegraphics[width=\linewidth]{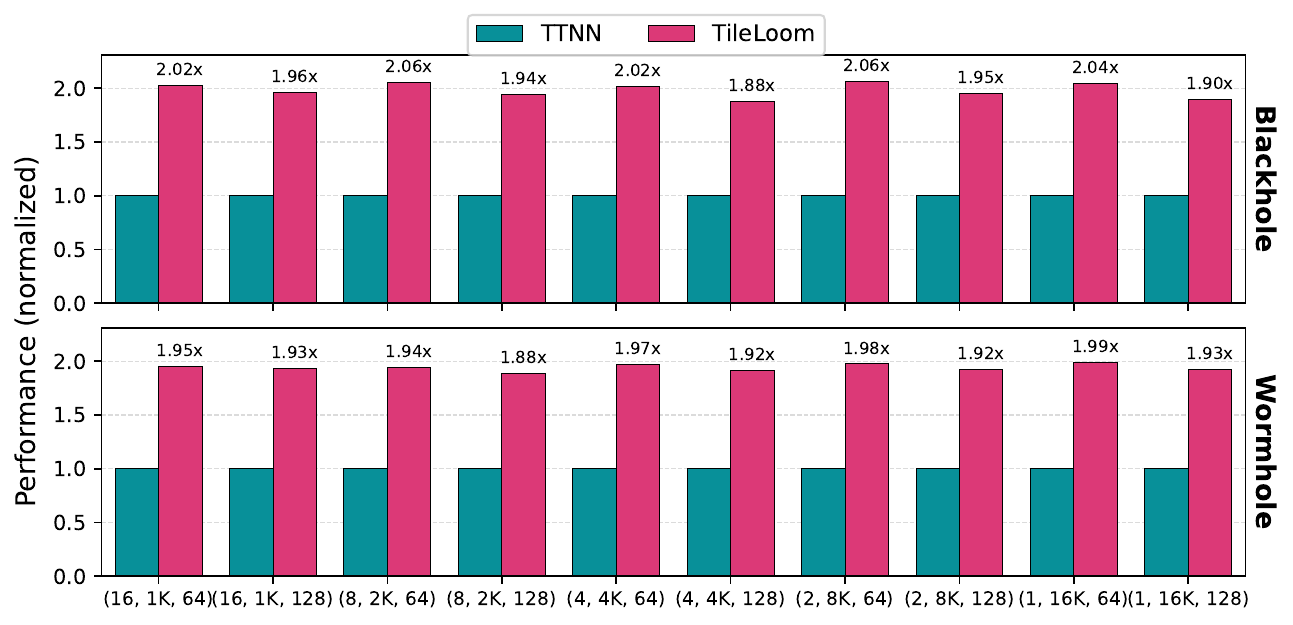}
    \caption{Relative Performance with {\loom} vs. TTNN on FlashAttention with parameters (batch size B, sequence length L, number of heads N).}
    \label{fig:fa}
\end{figure}

We evaluate {\loom} on non-causal FlashAttention against the native TTNN implementation. The non-causal variant exposes more dataflow-optimization opportunities than the causal variant, making it a useful stress test for {\loom}'s spatial mapping strategy. We vary the number of attention heads between 64 and 128, sweep sequence length from 1024 to 16384, and adjust batch size to fit within DRAM capacity.
As shown in Figure~\ref{fig:fa}, {\loom} consistently achieves substantial speedups over TTNN, with 1.88--2.06$\times$ improvement in nearly every configuration. The gains come from exploiting reuse in the attention operands: {\loom} places tiles so that key tiles are reused on chip across multiple query and value tiles, reducing DRAM traffic compared with TTNN's default mapping, which repeatedly reloads these operands from DRAM.

\subsubsection{FlashDecode}

Flash Decode is algorithmically a special case of FlashAttention with query length one, but this changes the hardware mapping problem significantly. In full FlashAttention, {\loom} can exploit parallelism across query blocks and heads, while choosing mappings that improve key/value tile reuse. In Flash Decode, the query dimension no longer offers spatial parallelism, so parallelism mainly comes from the batch dimension and from splitting the key/value reduction across cores.

This yields a much smaller dataflow-planning space: there are fewer ways to distribute independent output tiles, and the main optimization is partitioning the key/value sequence and orchestrating the cross-core gather-reduce. Thus, this benchmark stresses {\loom}'s generated block-level code and reduction orchestration more than its ability to choose among diverse dataflow mappings.

\begin{figure}[h]
    \centering
    \includegraphics[width=\linewidth]{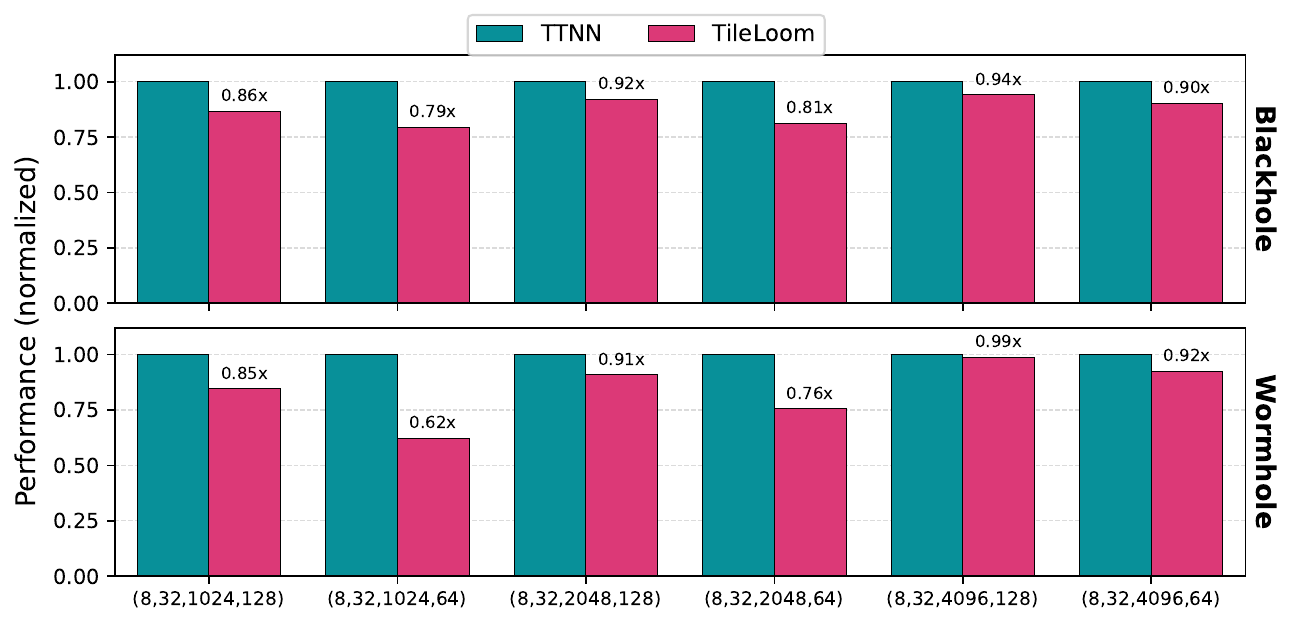}
    \caption{Relative performance of {\loom} compared with TTNN on Flash Decode with parameters (batch size B, head dimension H, sequence length L, hidden dimension D).}
    \label{fig:fd}
\end{figure}

As shown in Figure~\ref{fig:fd}, {\loom} does not outperform the TTNN Flash Decode baseline, which is expected because TTNN uses scheduling and reduction optimizations specifically tuned for this operator. In contrast, {\loom} generates its implementation through a general compiler stack starting from tile-level kernel code. Even so, {\loom} achieves about 85\% of TTNN performance on average, showing that it can produce competitive code even with limited dataflow optimization opportunity and a highly specialized vendor baseline.

\subsubsection{Mamba Chunk Scan}

Mamba Chunk Scan is a linear-attention kernel with substantial intra-block computation and nontrivial data movement. Since TTNN does not provide a fused implementation, we build the TTNN baseline by composing existing TTNN operations, yielding an unfused implementation. In contrast, {\loom} uses a fused tile-level Helion kernel.

\begin{figure}[h]
    \centering
    \includegraphics[width=\linewidth]{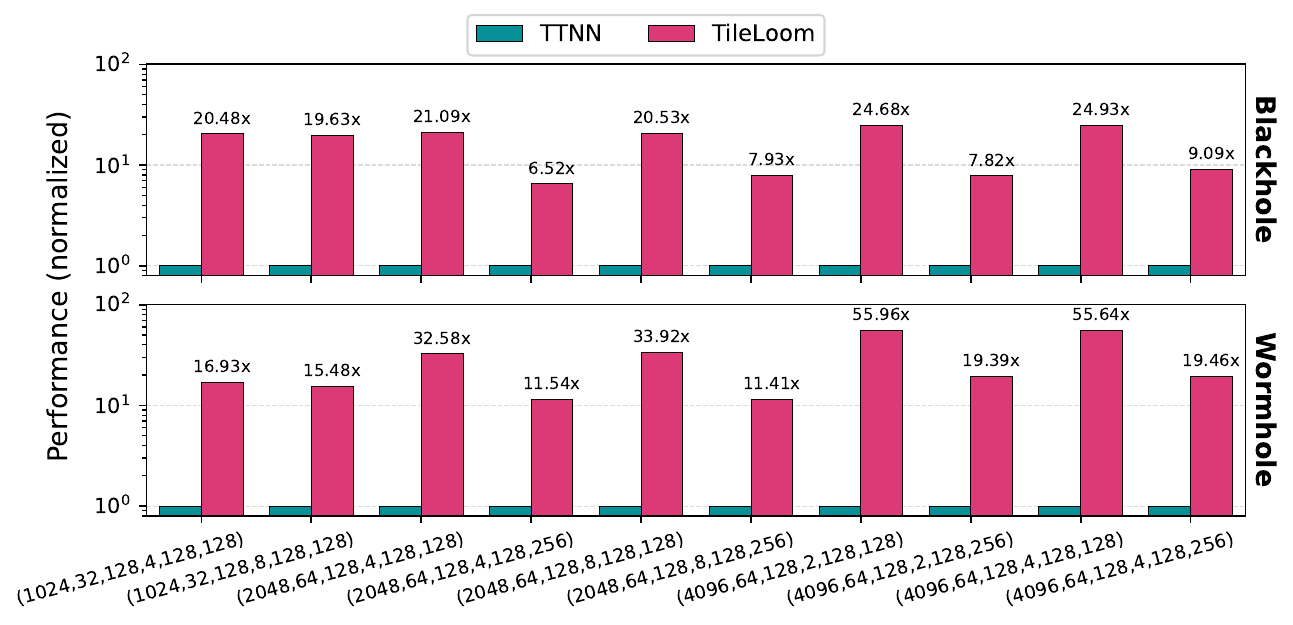}
    \caption{Relative performance of {\loom} vs. an unfused TTNN implementation on Mamba Chunk Scan with parameters sequence length L, number of heads N, head dimension H, number of groups G, hidden dimension D, and chunk size C.}
    \label{fig:mamba}
\end{figure}

This comparison shows a practical advantage of supporting tile-level kernel languages: fused kernels are often easier to obtain than highly optimized vendor-library implementations. Thus, {\loom} can generate efficient fused code even when the vendor library only exposes lower-level building blocks. As shown in Figure~\ref{fig:mamba}, {\loom} achieves 10x--55x speedup over the unfused TTNN baseline, reflecting both kernel fusion and {\loom}'s dataflow planning.

This large gap is typical between fused and unfused kernels, and similar trends appear in Tenstorrent's official \href{https://github.com/tenstorrent/tt-metal/tree/main/tech_reports}{technical reports}. It is especially pronounced on dataflow architectures, which rely on on-chip NoC bandwidth for reuse and often have less off-chip bandwidth than HBM-based GPUs. Fusion and dataflow planning are therefore critical: without them, intermediate tensors are repeatedly written to and read from off-chip memory, causing severe slowdowns.

\subsubsection{GEMM}

We evaluate GEMM on over $M$, $K$, and $N$, ranging from 256 to 16384. GEMM is a key primitive, and the TTNN vendor implementation is already highly optimized, achieving about 70\% of peak hardware throughput on average. Against this strong baseline, Table~\ref{tab:all-perf} shows that {\loom} delivers comparable performance using tile-level kernels while automatically generating low-level implementations.

\begin{figure*}
    \centering
    \includegraphics[width=.99\linewidth]{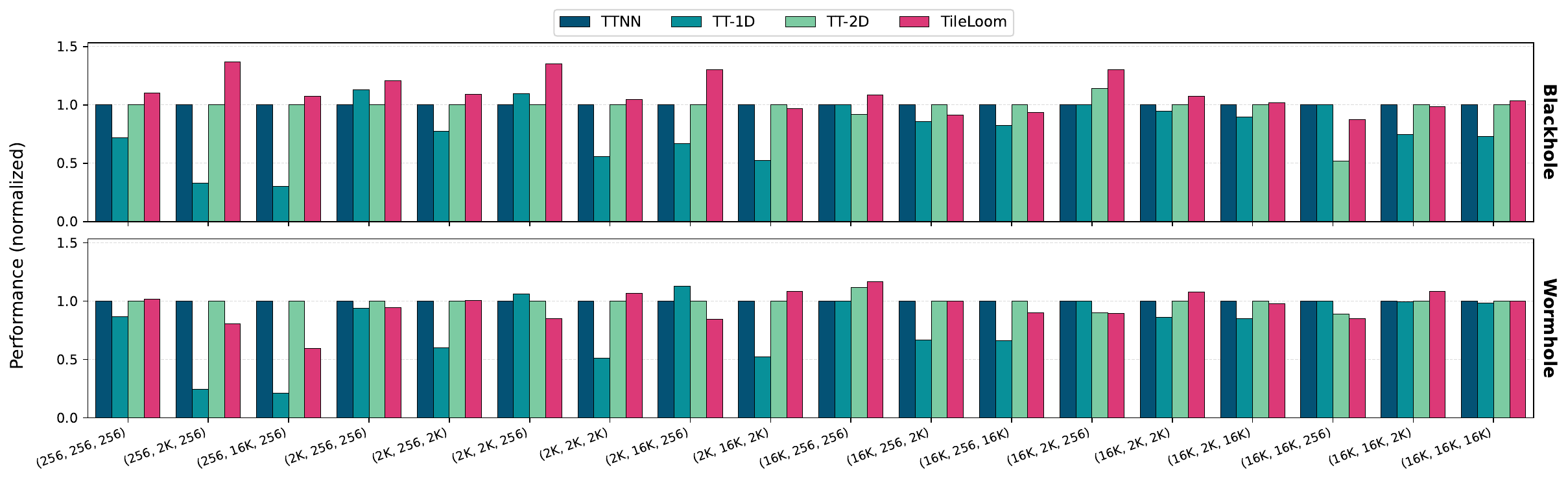}
    \caption{GEMM performance of {\loom} compared with TTNN. TT-1D and TT-2D are included as reference dataflow templates.}
    \label{fig:gemm}
\end{figure*}

For a more interpretable comparison, Figure~\ref{fig:gemm} also reports two TTNN dataflow templates, TT-1D and TT-2D. TT-1D uses a 1D broadcast pattern: each core loads the smaller input matrix from global memory, while the other input is broadcast across the entire array. TT-2D uses a 2D broadcast pattern, streaming the two inputs across the mesh from the top and left. TTNN selects between these templates using a shape-dependent heuristic, with block size chosen by a separate strategy.

Figure~\ref{fig:gemm} shows that {\loom} remains competitive with TTNN across the full GEMM sweep. This is notable because TTNN uses manually optimized library kernels, whereas {\loom} starts from tile-level kernels and searches the mapping space automatically. The TT-1D and TT-2D results show that fixed dataflows work well for regular GEMM shapes, where $M$, $K$, and $N$ are similar, but can degrade on irregular shapes where the best mapping depends more sensitively on dimensions and hardware balance.

\subsubsection{Shape Sensitivity on Irregular GEMM}

\begin{figure}[h]
    \centering
    \includegraphics[width=\linewidth]{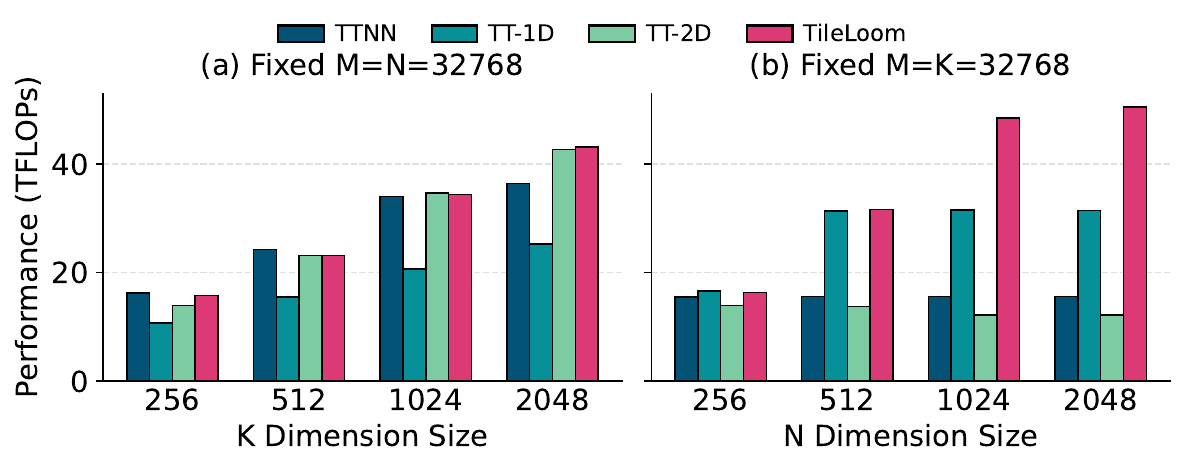}
    \caption{Performance comparison for GEMM under irregular input shapes.}
    \label{fig:gemm_irregular}
\end{figure}

To study shape sensitivity, we evaluate two irregular GEMM families. First, we fix $M = N = 32768$ and vary $K$ from 256 to 2048. Second, we fix $M = K = 32768$ and vary $N$ over the same range. Figure~\ref{fig:gemm_irregular} shows the results.

When varying $K$ (Figure~\ref{fig:gemm_irregular}a), {\loom} and TTNN follow trends similar to the 1D and 2D baselines. This is expected because $K$ is mapped sequentially within each core, so changing it mainly affects intra-core compute cost and leaves limited room for inter-core dataflow optimization.

In contrast, varying $N$ (Figure~\ref{fig:gemm_irregular}b) significantly changes the preferred dataflow. As $N$ approaches $M$, the workload becomes more balanced across rows and columns, making 2D-like broadcast more attractive due to reuse along both mesh dimensions. When $N$ is much smaller than $M$, 1D-like strategies are more effective, as reflected by the TT-1D and TT-2D results.

The transition between 1D- and 2D-favorable regimes depends on block size, compute-to-memory ratio, NoC overhead, and other hardware factors. This sensitivity motivates {\loom}'s cost-model-guided search, which accounts for spatial reuse, communication volume, and architectural constraints instead of relying on fixed heuristics.

These irregular-shape results also reveal a limitation of TTNN's heuristic selection: TTNN does not always choose the faster of its two templates. For example, in small-$N$, large-$M$ cases, TTNN selects TT-2D even though TT-1D is faster, such as $M,K,N = 32K,32K,512/1024$. Our inspection suggests this comes from fixed thresholds over ratios such as $M/N$. Near these thresholds, the best strategy can also depend on tile size, bandwidth, and NoC cost, but the heuristic still makes a hard ratio-based decision.

In contrast, {\loom} searches beyond the two TTNN templates, exploring how logical dimensions map to spatial dimensions, where memory operations are hoisted in the loop nest, and which block sizes to use. This broader search allows {\loom} to find mappings that outperform both TTNN templates, as shown in Figure~\ref{fig:gemm}.

\subsubsection{Wormhole vs. Blackhole}

{\loom} generally achieves larger relative gains on Blackhole than on Wormhole. This is because Blackhole increases compute throughput more than off-chip bandwidth: peak FP16 throughput improves by 2.53$\times$, while off-chip bandwidth improves by only 1.78$\times$ (Table~\ref{tab:tt-specs}). This shift makes kernels more likely to become memory-bound, increasing the value of {\loom}'s dataflow optimizations, which improve spatial reuse and reduce off-chip traffic.

GEMM shows this trend most clearly. On Wormhole, {\loom} reaches 0.95$\times$ TTNN performance. Profiling shows that GEMM is often compute-bound on Wormhole, so reducing off-chip traffic has limited end-to-end benefit. In addition, {\loom} starts from tile-level kernels and does not yet include all microarchitecture-specific optimizations used in TTNN's hand-tuned GEMM library, explaining the remaining 5\% gap.

On Blackhole, the same GEMM evaluation improves to 1.10$\times$ over TTNN. Blackhole's higher compute-to-bandwidth ratio makes GEMM more sensitive to memory traffic and placement, allowing {\loom}'s dataflow planning to provide greater benefit. This shows that while {\loom} may not always beat a vendor GEMM library on a compute-limited device, its architectural modeling and spatial-reuse optimizations become more valuable as hardware shifts toward higher compute density.

\subsection{Ablation Studies}
\subsubsection{Effect of Spatial Reuse}

We quantify how much of {\loom}'s GEMM performance comes from spatial reuse by disabling the spatial-reuse pass and forcing all operands to be loaded from DRAM. Table~\ref{table:ablation_spatial} reports absolute performance with and without this optimization.
Spatial reuse gives the largest gains on smaller GEMMs, where DRAM traffic is a major bottleneck. As size grows, the benefit decreases, consistent with the roofline model~\cite{roofline}: GEMM arithmetic intensity increases with matrix size, making larger problems more compute-bound. Once performance is limited by peak compute, reducing DRAM traffic has diminishing runtime impact.\footnote{Across all GEMM measurements, including configurations not shown here, effective throughput stabilizes around 45~TOP/s, indicating that many configurations operate near the compute roof.}
Still, spatial reuse substantially reduces memory pressure: across these GEMM configurations, it cuts DRAM accesses by 70\% on average. This does not always translate to proportional speedup in compute-bound regimes, but can improve memory headroom, reduce power, and leave more bandwidth for concurrent workloads.

\begin{table}[h]
\caption{GEMM performance (TFLOP/s) on Wormhole with ({\loom}) and without (DRAM only) spatial reuse.}
\label{table:ablation_spatial}
\centering
%\resizebox{.9\linewidth}{!}{
\begin{tabular}{lccc}
\hline
Configuration & DRAM only & {\loom} & Speedup \\ \hline
M=K=N=1024 & 11.15 & 23.70 & 2.12$\times$ \\
M=K=N=2048 & 16.77 & 33.96 & 2.03$\times$ \\
M=K=N=4096 & 22.96 & 40.44 & 1.76$\times$ \\
M=K=N=5120 & 29.19 & 41.61 & 1.42$\times$ \\
M=K=N=6144 & 28.22 & 42.47 & 1.51$\times$ \\ \hline
\end{tabular}
%}
\end{table}

\subsubsection{Effect of Temporal Reuse}

\begin{figure}[htbp]
    \centering
    \includegraphics[width=.8\linewidth]{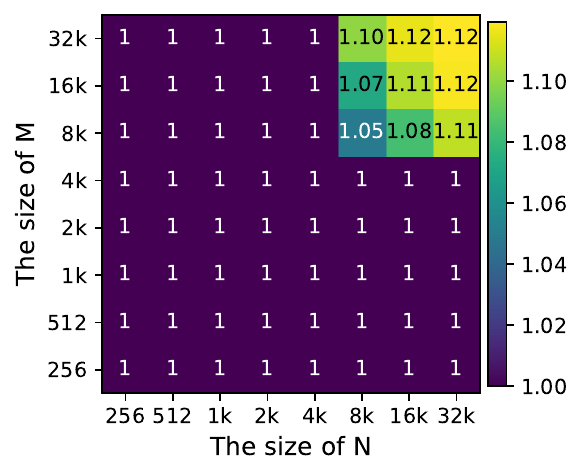}
    \caption{Normalized GEMM performance on Wormhole with and without temporal reuse.}
    \label{fig:gemm_temporal_reuse}
\end{figure}

Figure~\ref{fig:gemm_temporal_reuse} shows the impact of temporal reuse across GEMM shapes. Temporal reuse buffers tiles locally so the same $A$ or $B$ tiles are reused across iterations instead of repeatedly loaded from DRAM.

We compare {\loom} with and without temporal reuse over different $M$ and $N$ values. As with spatial reuse, this optimization is most useful in memory-bound regimes, so we decrease $K$ as $M$ and $N$ grow to keep the configurations memory-bound. In this setting, temporal reuse yields speedups up to 1.12$\times$. The benefit increases with $M$ and $N$, since larger dimensions create more opportunities to reuse the same $A$ or $B$ tiles. Thus, temporal reuse is most helpful when $M$ or $N$ is large and $K$ is small. When it provides little benefit, {\loom}'s performance model deprioritizes those mappings, resulting in the same selected mapping and performance as the baseline without temporal reuse.

\subsubsection{Accuracy of {\loom}'s Performance Model Alone}

\begin{figure}[htbp]
    \centering
    \includegraphics[width=\linewidth]{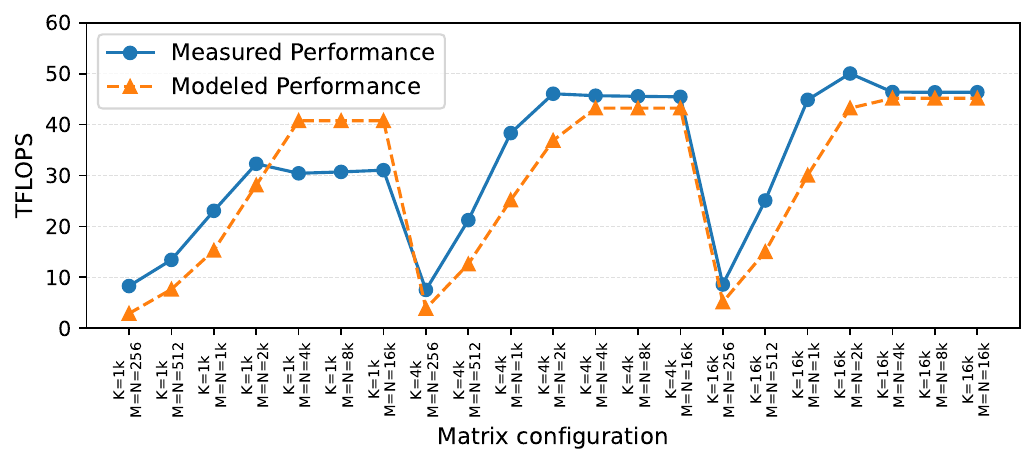}
    \caption{Validation of {\loom}'s performance model against measured GEMM performance.}
    \label{fig:validate_perf_model}
\end{figure}

We validate {\loom}'s performance model on Wormhole by comparing predicted throughput with measured GEMM performance across a wide range of $(M,N,K)$ configurations. Figure~\ref{fig:validate_perf_model} plots both estimates and measurements.

The predictions differ from measurements by 17\% in geometric mean. However, the goal is not cycle accuracy, but reliable ranking and trend prediction, especially for transitions between memory- and compute-bound regimes as described by the roofline model~\cite{roofline}. As shown in Figure~\ref{fig:validate_perf_model}, the model captures these transitions well: it identifies when configurations become compute-bound and reflects the relative performance changes across shapes.

The end-to-end results in Table~\ref{tab:all-perf} and ~\ref{table:ablation_top_k} further show that this error has limited impact on mapping selection. The model is accurate enough to rank candidates effectively, leaving the final optional profiling stage to choose among a small set of high-quality mappings.
% \subsubsection{Accuracy of {\loom}'s Performance Model Alone}
% We validate {\loom}'s performance model on Wormhole device by comparing its predicted throughput against measured hardware performance for GEMM over a wide range of $(M, N, K)$ configurations. Figure~\ref{fig:validate_perf_model} plots both the model’s estimates and the actual measured performance.

% On average, the predicted performance differs from measurements by 17\% in geometric mean. Our goal, however, is not a cycle-accurate predictor, but a model that reliably captures relative performance trends, especially transitions between memory-bound and compute-bound regimes as described by the roofline model~\cite{roofline}. As shown in Figure~\ref{fig:validate_perf_model}, the model tracks these transitions well: it consistently identifies when a configuration becomes compute-bound, and it reflects the relative magnitude of performance changes across shapes.

% Most importantly, the end-to-end results in Table~\ref{table:ablation_top_k} show that this level of error has limited impact on {\loom}'s ability to choose good mappings. The model is accurate enough to guide dataflow selection and rank candidates so that the final profiling stage only needs to discriminate among a small, high-quality set.

\subsubsection{Effect of top-$k$ tuning and different topologies}
\label{sec:ablation_topk}

As discussed in Section~\ref{subsec:perf-model}, {\loom} can optionally profile the top-$k$ candidates generated by the compiler and select the best one. Thus, $k$ trades compilation cost for final performance. We vary $k$ from 1 to 5 and report geometric-mean normalized performance on Wormhole, along with compile time across hardware topologies, in Table~\ref{table:ablation_top_k}. Here, top-1 means fully static compilation with only the best-predicted mapping, while larger $k$ values add more profiled candidates.

We evaluate three topologies to test sensitivity to available parallelism and communication structure. The $8 \times 8$ mesh uses the full Wormhole core array, while the $4 \times 8$ mesh and $1 \times 8$ ring model smaller subsets. These reduced topologies change the balance between compute, NoC communication, and off-chip bandwidth, stressing whether mappings predicted for the full mesh remain effective under more constrained layouts.

Overall, $k$ has a modest effect on performance. On the $8 \times 8$ mesh, top-5 outperforms top-1 by 7\%, with most of the gain already achieved at top-2, which improves performance by 4.7\%. The remaining gap mainly comes from occasional model misrankings, where the predicted-best mapping is slightly worse than another candidate. In practice, a small $k$, such as 2 or 3, is usually enough to include the best mapping while keeping compile time moderate.

Similar trends hold on smaller topologies. The $1 \times 8$ ring is least sensitive to $k$ because its restricted layout leaves fewer strong mappings to distinguish. The $4 \times 8$ and $8 \times 8$ meshes expose larger search spaces, so profiling a few extra candidates helps more. However, gains saturate quickly: beyond top-3, performance improves only marginally while compile time grows nearly linearly. This suggests that {\loom}'s static model effectively narrows the search to a small set of promising mappings, and a small profiling budget recovers most available performance across topology scales.

\begin{table}[h]
    \caption{Geometric-mean normalized performance (relative to TTNN) and compile time (seconds) of {\loom} under different top-$k$ profiling settings. Top-1 corresponds to fully static compilation without additional profiling.}
    \label{table:ablation_top_k}
    \centering
%    \resizebox{.85\linewidth}{!}{
    \begin{tabular}{c|ccc}
    \hline
    & $1\times8$ Ring & $4\times8$ Mesh & $8\times8$ Mesh \\ \hline
    top-1 & -1.3\% (6.36)  & -5.4\% (5.53)  & -6.1\% (5.75)  \\
    top-2 & -1.3\% (12.70) & -0.8\% (11.29) & -1.4\% (11.17) \\
    top-3 & -1.2\% (19.07) & +0.5\% (17.16) & -0.5\% (16.65) \\
    top-4 & -0.4\% (25.47) & +1.1\% (23.03) & +0.4\% (22.17) \\
    top-5 & -0.3\% (31.82) & +1.3\% (28.91) & +0.9\% (27.66) \\ \hline
    \end{tabular}
%    }
\end{table}

\section{Related Works}
\noindent\textbf{Hardware modeling and co-design.}
There exist works on modeling and co-design of spatial accelerators: CGRAs, spatial FPGAs, and systolic arrays, where the computes and control are at a finer granularity than {\loom}. Languages and frameworks such as Spatial~\cite{spatial}, Plasticine~\cite{plasticine}, T2S~\cite{T2S}, Halide(-to-Hardware)~\cite{halide}, and HeteroCL~\cite{heterocl} compile loop nests or functional pipelines into arrays of processing elements (PEs) and local memories, often co-designing the overlay itself. Polyhedral and tensor-centric systems such as AutoSA~\cite{autosa}, TensorLib~\cite{tensorlib}, and Rubick~\cite{rubick1,rubick2} similarly start from affine loop nests or tensor expressions and derive space–time mappings that synthesize systolic or tensor arrays, buffers, and controllers. Co-design tools like Timeloop/Accelergy~\cite{timeloop} and AMOS~\cite{amos} explore dataflows, tilings, and memory hierarchies for DNN accelerators, while MAESTRO/MAERI~\cite{maestro,maeri} analytically model buffer and network usage for specialized systolic-style designs. In all of these systems, the modeling unit is a PE, buffer, or loop level in the memory hierarchy, and the goal is to search or synthesize hardware at that granularity. {\loom}, in contrast, assumes intra-core microarchitecture and local mapping are fixed, treats each core as the atomic unit, and models only the multi-level memory system and NoC \emph{between} cores to decide how tile instances are distributed across space and time.

\noindent\textbf{Software mapping and compilation for spatial architectures.}
On the software side, classical CGRA and FPGA flows treat mapping as a place-and-route problem on a PE-level dataflow graph: frameworks such as DSAGEN~\cite{dsagen}, ML-oriented CGRA compilers like ML-CGRA and MLIR-to-CGRA~\cite{mlcgra,mlircgra}, and architecture-agnostic mappers such as Morpher~\cite{morpher} and CaSMap~\cite{casmap} perform placement, routing, and modulo scheduling onto a fixed fabric. Loop- and polyhedral-based tools including Timeloop/Accelergy~\cite{timeloop}, and MAESTRO/MAERI~\cite{maestro,maeri} reuse loop schedules or space–time mappings as the scheduling representation but are primarily design-space exploration tools: they evaluate mappings analytically. Research compilers for specific spatial architecture, such as AMOS~\cite{amos}, LISA~\cite{lisa}, and system-level compilers for wafer-scale fabrics~\cite{cerebras_compiler_mach,cerebras_compiler_osdi}, do generate per-core programs and communication schedules, but are typically tailored to a particular architecture family with baked-in mapping heuristics.

%{\loom} combines ideas from these lines of work while targeting existing spatial-dataflow accelerators. It uses tile-wise kernels (e.g., from a Triton-like DSL) as its software IR, an explicit architecture description akin to loop- and tensor-based cost models, and a mapping search integrated into an MLIR-based compiler pipeline. The same spatial–temporal mapping representation drives both the analytic cost model and the synthesis of NoC transfers, synchronization, and memory operations, so that selected mappings directly produce executable per-core programs on different spatial-dataflow architectures instead of only guiding hardware design or a single custom stack.

\section{Conclusion}
%TileLoom demonstrates that compiler-driven mapping can deliver vendor-level performance on spatial dataflow accelerators while dramatically reducing manual effort. Starting from high-level, tile-centric kernels, TileLoom automatically selects spatial–temporal mappings, data-reuse schemes, and communication patterns using a dataflow dialect that uniformly captures cores, memories, and interconnect. On workloads such as GEMM and FlashAttention, this approach achieves competitive or even better performance than vendor libraries with minimal human intervention, turning hand-engineered kernels into the output of a reusable compilation pipeline and enabling a broader set of developers to author new kernels. At the same time, TileLoom’s hardware abstractions make compiler passes largely reusable across accelerators, easing support for diverse spatial dataflow architectures and providing a natural substrate for design-space exploration of future architecture designs.

% Looking ahead, we plan to extend TileLoom along both its front-end and back-end. On the front-end side, we aim to support additional DSLs and IRs, such as Helion, to broaden the range of kernels and programming styles the system can ingest. On the back-end side, we plan to target more spatial dataflow architectures, including emerging engines such as AMD’s NPU and AIE families, further testing the portability of the dataflow dialect and mapping algorithms across diverse accelerator designs.

TileLoom demonstrates that compiler-driven mapping can deliver competitive performance on spatial dataflow accelerators while substantially reducing the need for hand-written, hardware-specific kernels. Starting from high-level tile-centric kernels, TileLoom automatically selects spatial and temporal mappings, data-reuse strategies, and communication patterns using a dataflow dialect that uniformly represents cores, memories, and interconnect.

Across Tenstorrent Wormhole and Blackhole, TileLoom achieves strong results on FlashAttention, Mamba Chunk Scan, GEMM, and Flash Decode. It outperforms TTNN when dataflow planning and fusion expose substantial opportunities, and remains competitive with highly optimized vendor implementations when the optimization space is more constrained. These results show that many decisions traditionally embedded in hand-engineered kernels can instead be handled by a reusable compiler stack, making optimized kernel development more accessible and providing a foundation for future spatial dataflow architectures.

\section*{Acknowledgment}
This work is partially supported by the Advanced Research and Technology Innovation Centre (ARTIC), the National University of Singapore under Grant AFP-RP6, and by National Research Foundation, Singapore, under its Competitive Research Program Award NRF-CRP23-2019-0003 and
the Ministry of Education, Singapore, under Tier 3 grant MOE-MOET32024-0003.

\clearpage
%-------------------------------------------------------------------------------
\bibliographystyle{plain}
\bibliography{ref}

%%%%%%%%%%%%%%%%%%%%%%%%%%%%%%%%%%%%%%%%%%%%%%%%%%%%%%%%%%%%%%%%%%%%%%%%%%%%%%%%
\end{document}